\documentclass{elsart}
\usepackage{natbib}
\usepackage[dvips]{graphics, graphicx, epsfig}
\usepackage{caption}
\usepackage{amsmath}

\begin{document}
\runauthor{F. Wauters et al.}
\begin{frontmatter}
\title{Performance of silicon PIN photodiodes at low temperatures and in high magnetic fields}

\author[IKS]{F. Wauters\corauthref{cor1}},
\ead {Frederik.Wauters@fys.kuleuven.be} \corauth[cor1] {Tel.:
+32-16-327512; fax: +32-16-327985.}
\author[IKS]{I.S. Kraev},
\author[IKS]{M. Tandecki},
\author[IKS]{E. Traykov},
\author[IKS]{S. Van Gorp},
\author[NPI]{D. Z\'akouck\'y},
\author[IKS]{N. Severijns}

\address[IKS]{K.U. Leuven, Instituut voor Kern- en Stralingsfysica, Celestijnenlaan 200D, B-3001 Leuven, Belgium}
\address[NPI]{Nuclear Physics Institute, ASCR, 250 68 \v{R}e\v{z}, Czech Republic}

\begin{abstract}
The performance of a Si PIN diode (type Hamamatsu S3590-06) as an
energy sensitive detector operating at cryogenic temperatures
($\sim$10 K) and in magnetic fields up to 11 T was investigated,
using a $^{207}$Bi conversion electron source. It was found that
the detector still performs well under these conditions, with
small changes in the response function being observed in high
magnetic fields, e.g. a 30\% to 50\% decrease in energy
resolution. A GEANT4 Monte Carlo simulation showed that the
observed effects are mainly due to the modified trajectories of
the electrons due to the influence of the magnetic field, which
changes the scattering conditions, rather then to intrinsic
changes of the performance of the detector itself.
\end{abstract}


\end{frontmatter}

\section{Introduction}

Si PIN photodiodes, originally designed to detect photons, have
over the years found more and more applications in nuclear and
particle physics. Nowadays they are used in a wide variety of
experiments for the detection of X-rays \cite{Inoue:2002},
electrons \cite{Schuurmans:2000,Wustling:2002,Kraev:2005}, or
$\alpha$-particles
\cite{Krause:1998,Schuurmans:1999,Severijns:2005}, but they have
up to now always been used in rather low magnetic fields.
\\
Solid state detectors like avalanche photodiodes and silicon drift
detectors have already been tested in high magnetic fields up to
respectively 7.9 T \cite{Marler:2000} and 4.7 T \cite{Pandey:1996}
to be used in tracking devices at collider experiments. However,
these detectors were not used for spectroscopic purposes. With
Penning ion traps becoming more widespread in nuclear physics
\cite{Kluge:2002}, and now also being used for in-trap and
trap-assisted spectroscopic measurements (e.g.
\cite{Kurpeta:2007}), an increasing need for solid state state
detectors that can operate in even higher magnetic fields (up to
11 T) has emerged. For $\alpha$-particles and low energy electrons
a Si PIN diode is a good option provided it performs well in such
extreme conditions.
\\
Here we present the results of measurements to test the
performance of a Si PIN diode at a temperature of $\sim$ 10 K and
in magnetic fields up to 11 T using conversion electrons from a
$^{207}$Bi source. A series of GEANT4 \cite{Agostinelli:2003}
simulations was performed as well to get a better understanding of
these results.

\section{Experimental setup}

The performance of a Si PIN diode at about 10 K and in high
magnetic fields was measured using a Brute Force Low Temperature
Nuclear Orientation setup \cite{Kraev:2005}. The apparatus is
equipped with a superconducting solenoid providing a magnetic
field up to 17 T.
\\The detector was a Si PIN photodiode produced by Hamamatsu Photonics
(type S3590-06). This is a 500 $\mu$m thick fully depleted
windowless PIN diode with a sensitive area of 10 x 10 mm$^2$. It
allows to fully stop electrons with energies up to $~$350 keV,
while the sensitivity for $\gamma$ radiation is still very low.
The insensitive SiO$_{2}$ surface dead layer has a thickness of
270 nm \cite{Akimoto:2006} which is thin enough to allow to use
this PIN diode also for the detection of $\alpha$ particles.
\\The housing consists only of ceramic material and non-magnetic
metals, rendering this detector well-suited for operation at
cryogenic temperatures and in magnetic fields. The detector was
previously already commissioned and used in experiments at a
temperature of $\sim$10 K and in a magnetic field of 0.6 T,
showing good behaviour in these conditions \cite{Kraev:2005}.
\\The electron source used was a commercially available $^{207}$Bi
conversion electron source. The decay of this isotope proceeds
mainly via three $\gamma$ lines with energies of 539.7 keV, 1063.7
keV and 1770.2 keV, respectively, each of them being converted for
a couple of percent yielding several conversion electron lines.
Only the conversion electrons of the two lowest energetic
transitions were used here because of the thickness of the
detector. The $^{207}$Bi activity is sandwiched between two 2.4
mg/cm$^2$ thick titanium windows supported by an aluminium ring. A
GEANT4 simulation showed that energy loss in these windows shifts
the full energy peak of the 481.7 keV and 975.7 keV K conversion
electrons by 2.3 keV respectively 2.0 keV to lower energies and
causes an additional energy spread of 1.9 keV and 1.7 keV,
respectively.
\\The source was placed at a distance of 8 mm from the Si PIN diode
with a 7 mm long, 10 mm diameter Al collimator in between and a 1
mm thick, 9 mm diameter Cu collimator on top of the detector. The
Cu collimator served to limit the active area of the detector as
it is known that the charge collection efficiency for this kind of
detectors degrades at the edges of the sensitive surface
\cite{Simon:2005}. The system was installed in the bore tube of
the superconducting magnet as shown in fig.~\ref{fig:experimental
setup}. The positioning is such that the magnetic field lines are
perpendicular to the surface of the detector. The signal is led
through the 4 Kelvin, 77 Kelvin and room temperature radiation
shields to a charge collecting pre-amplifier (Canberra 2001)
outside the cryostat. It was further processed by standard NIM
electronics and a PC based data acquisition system.

\begin{figure}[ht]
\centering
\includegraphics[width=\columnwidth]{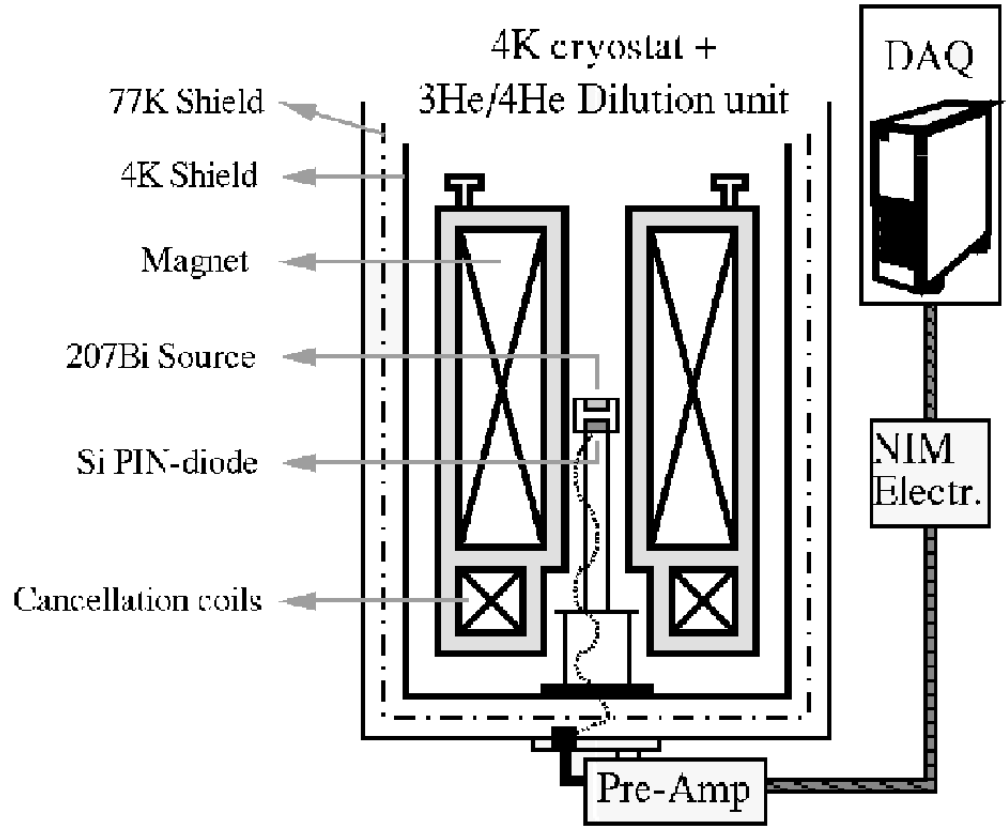}
\caption {Sketch of the experimental setup. The Si PIN photodiode
and the $^{207}$Bi source are placed inside the bore tube of the
17 T superconducting magnet. The signal from the detector passes
through the 4 K, 77 K and room temperature radiation shields
before reaching the pre-amplifier. } \label{fig:experimental
setup}
\end{figure}

\section{Measurement and results}

\begin{figure}[ht]
\centering
\includegraphics[width=\columnwidth]{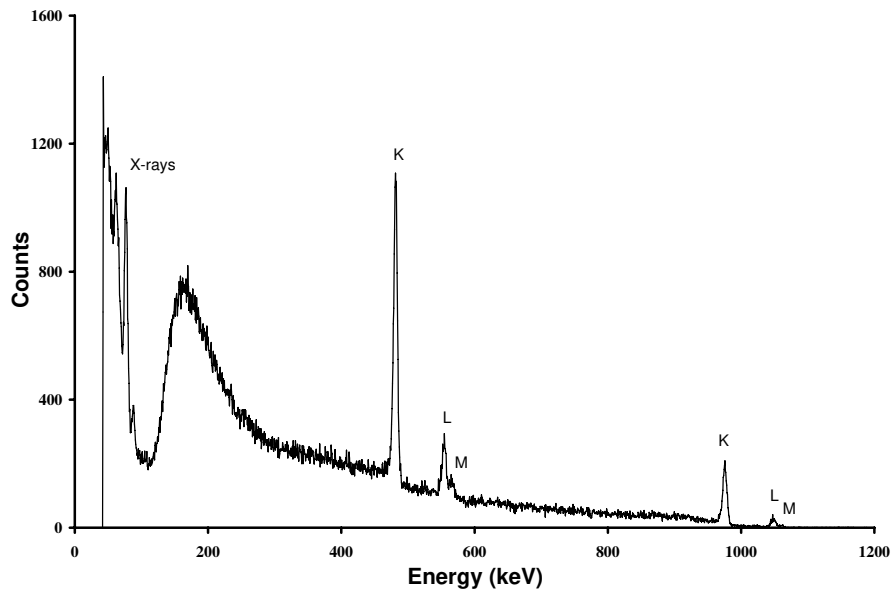}
\caption {Conversion electron spectrum of $^{207}$Bi measured with
the Si PIN diode operating at $\sim$10 K in 0 T. The reverse bias
was 140 V.} \label{fig:reference spectrum}
\end{figure}

\begin{figure}[ht]
\centering
\includegraphics[width=\columnwidth]{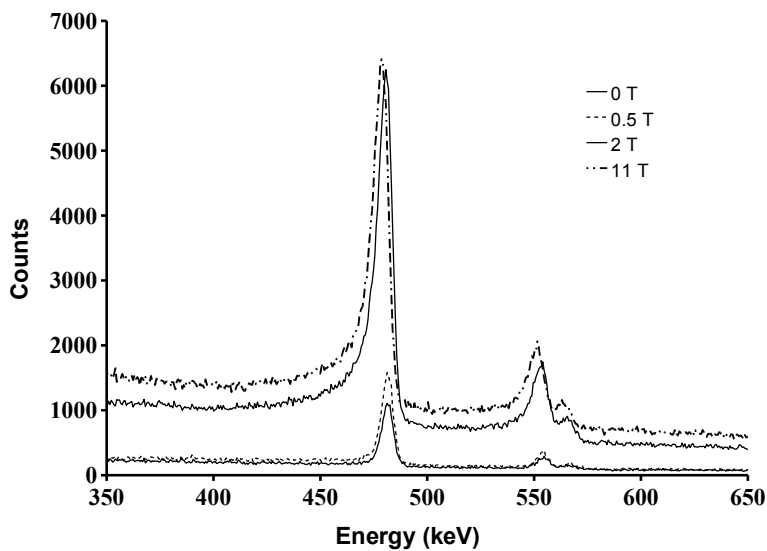}
\caption {Conversion electron spectra ($^{207}$Bi) obtained with
the Si PIN diode operating at $\sim$10 K and in magnetic fields of
0 T, 0.5 T, 2 T and 11 T. The most important change is the
counting rate due to the focussing of the electrons while, in
addition, the shape and the position of the full energy peaks also
change slightly (see text).} \label{fig:field spectra}
\end{figure}

First the detector was cooled to $\sim$10 K in the absence of a
magnetic field. In accordance with previous measurements
\cite{Kraev:2005} no significant difference between the $^{207}$Bi
spectra measured at room temperature and at cryogenic temperatures
was observed. Nevertheless it is well known that cooling solid
state detectors generally improves the energy resolution since the
leakage current of the detector significantly decreases when
cooling it below $0{^\circ}$ C \cite{Weinheimer:2006}. The fact
that the performance of the diode at room temperature and at
cryogenic temperatures was similar in our setup then indicates
that the major contribution to the noise is in this case not
coming from the detector itself but from the signal processing.
Indeed, the complexity of a cryogenic setup requires to lead the
unamplified signal from the detector through three radiation
shields (i.e at 4 K, at 77 K and at room temperature) to the
preamplifier. One has to try to avoid that this wiring acts too
much as a heat bridge between the room temperature parts and the 4
K cooled parts of the system, leading to a wire length of about 30
cm and increased pick-up of noise. The energy resolution (FWHM) of
the detector in this configuration was, both at room temperature
and at $\sim$10 K, 6.4(1) keV and 7.5(2) keV for the 482 keV and
976 keV conversion electron lines, respectively. A spectrum
obtained at $\sim$10 K and without magnetic field is shown in
figure \ref{fig:reference spectrum}. The two sets of conversion
lines are clearly visible above the background of scattered
electrons. The behavior of the conversion electron peaks at 482
keV and 976 keV was monitored throughout the different
measurements that were performed.
\\
During a first series of measurements the field at the position of
the detector was gradually increased from 0 T to 11 T. No major
changes in the performance of the detector were observed. Between
0 T and 2 T the count rate increased roughly by a factor of 10 due
to the focussing of the electrons by the magnetic field. This
resulted in a 2$\pi$ effective solid angle at about 2 T for the
482 keV electrons and at somewhat higher magnetic field also for
the 976 keV electrons. This is illustrated in
figure~\ref{fig:field spectra}. Note that the initial count rate
was sufficiently low so that the dead time, even at the highest
field values, did not get larger than a few percent.
In addition to the overall change in count rate a clear change in
the shape of the peaks as a function of the magnetic field is
observed; with the magnetic field increasing the resolution
worsens somewhat, while the peaks get a tail on the low energy
side and also move slightly to lower energies (i.e. about 2.5
keV). When the field was lowered back to 0 T the spectrum was
again identical to the one observed before the field was raised.
The effects observed in high magnetic fields were thus clearly due
to the presence of the field and no damage was caused to the
detector.
\\
In a second series of measurements data were taken with the
detector subjected to a field of 5 T for two full days. The
operation of the detector was observed to remain stable and no
change in the spectra was seen during this entire period, showing
that stable, long-term operation of the detector at low
temperatures and in high magnetic fields is possible.
\\
Finally, spectra were again recorded while the field was changed
in steps, now decreasing it from 11 T down to 0 T. Smaller steps
were taken in the region where the changes to the spectra seemed
to be largest, i.e. between 1 T and 5 T. No differences compared
to the first sweep, where the field was increased from zero field
to 11 T, was observed. This can be seen in figure~\ref{fig:field
resolution} where the energy resolution (FWHM) of the 482 keV and
the 976 keV conversion electron peaks as a function of the
magnetic field for the two field sweeps is shown. The resolution
of the pulser peak, present during the entire experiment, was 2.0
keV and stable within 0.1 keV.
\begin{figure}[ht]
\centering
\includegraphics[width=\columnwidth]{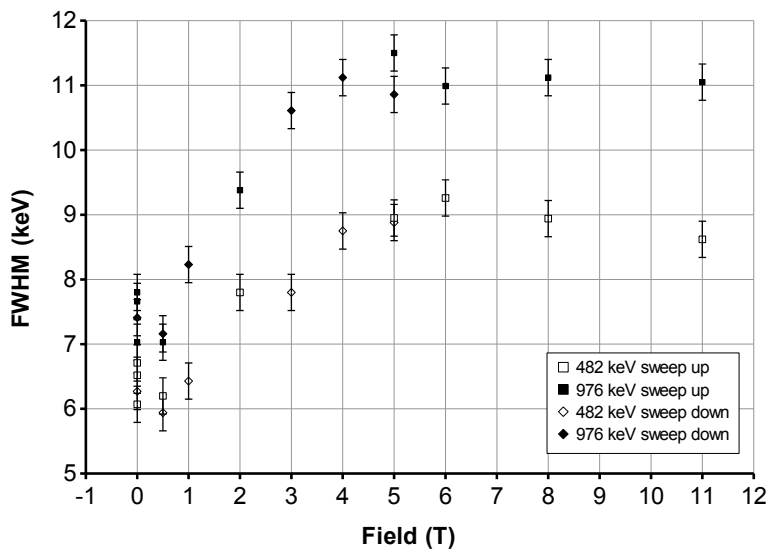}
\caption {Energy resolution of the full energy peaks of the 482
keV and 976 keV conversion electron lines of $^{207}$Bi observed
with a Si PIN diode operating in different magnetic fields and at
a temperature of $\sim$10 K.} \label{fig:field resolution}
\end{figure}

\section{Discussion and interpretation}

\begin{figure}[ht]
\centering
\includegraphics[width=\columnwidth]{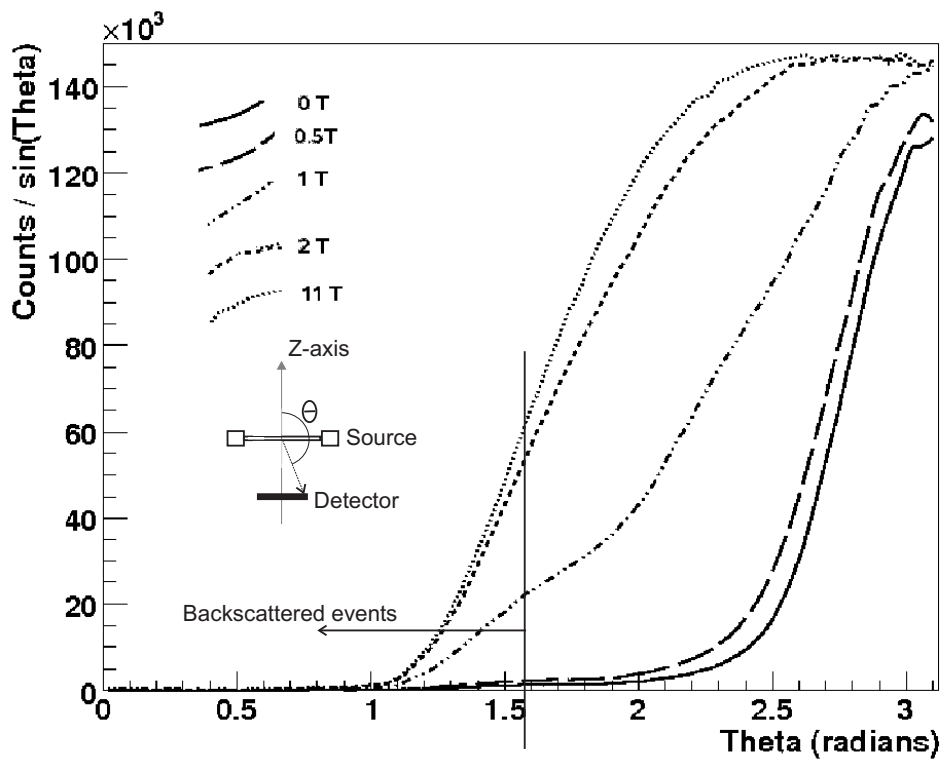}
\caption {The number of 500 keV electrons that end up in the full
energy peak as a function of their initial emission angle (defined
relative to the vertical axis and corrected for the solid angle)
for different applied magnetic fields. Already at 2 T about 46 \%
of the emitted electrons end up in the detector, while in a field
of 11 T about 50 \% do so. The events to the left of the vertical
solid line are backscattered electrons that still reach the
detector.} \label{fig:field solid angle}
\end{figure}

\begin{figure}[ht]
\centering
\includegraphics[width=\columnwidth]{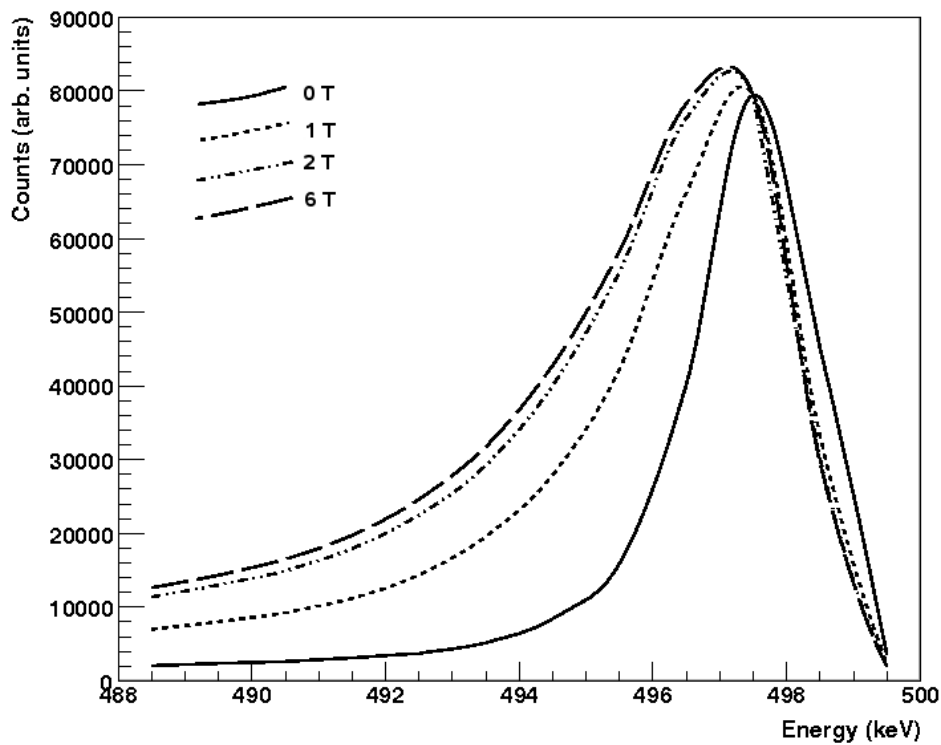}
\caption {The simulated detector response to 500 keV electrons for
applied magnetic fields up to 11 T. Note that the spectra are
rescaled so that the peak shapes can more easily be compared.}
\label{fig:field geant}
\end{figure}

Although the Si PIN diode tested here still works well in high
magnetic fields, a clear dependence of the spectrum shape on the
magnetic field strength was nevertheless observed (see
Fig.\ref{fig:field spectra}). This is important for the use of
this detector as a spectroscopic device.
\\
To investigate the origin of these changes in more detail a series
of GEANT4 (version 4.90) Monte Carlo simulations were performed.
The routine that was used was developed to track electrons and
reproduce experimental spectra for the real experiment
\cite{Kraev:2008}. A detailed geometrical description of both the
detector and the source, including also the piece supporting them,
as well as the collimators, were included. For ease of
interpretation mono-energetic electron lines of 500 keV and 1000
keV were included instead of implementing the full $^{207}$Bi
decay scheme. Further, different homogeneous magnetic fields,
comparable to the ones used in the experiment, were applied in
these simulations, while for every emitted electron both the
emission angle and the energy deposited in the sensitive area of
the detector were recorded. Figure \ref{fig:field solid angle}
shows, for different magnetic fields, the initial emission angle
relative to the vertical axis for the electrons that are detected
in the full energy peak. It is seen that from 2 T upward the
detector indeed has almost a 2 $\pi$ effective solid angle.
\\
\noindent In figure \ref{fig:field geant} the simulated detector
response to 500 keV mono-energetic electrons is shown for
different fields. It must be stressed that the Monte Carlo
simulation only registers the energy which the electrons deposit
in the detector and that effects of charge collection and signal
processing, which also contribute to the energy resolution in a
real experiment, are not included in the simulations.
\\
It is seen that the effects that were observed when the field was
increased, i.e. the fact that the full energy peak broadens, gets
a tail at the low energy side and also slightly shifts to lower
energies, are all reproduced by the simulation. This indicates
that these changes are mainly caused by the magnetic field that
modifies the trajectories of the $\beta$-particles so that they
deposit different energies in the sensitive volume of the
detector, rather than being due to an intrinsic change in the
performance of the detector itself. This can be rather easily
understood. With no field the majority of the detected
$\beta$~particles follow a straight trajectory from source to
detector and pass almost perpendicular through the source window
and through the detector front dead layer. In the presence of a
magnetic field they spiral from source to detector thereby
arriving at a rather shallow incidence angle on the surface of the
detector. This causes a longer and more varying path length in the
source window as well as in the front dead layer and in the
sensitive volume of the detector, leading to increased scattering
and backscattering as well as to a different energy deposition.
\\
Table \ref{tab:Simulation_Experiment} lists the experimentally
observed as well as the simulated change in energy resolution
(FWHM), together with the simulated effective solid angle, for
different magnetic fields and for electron energies of 500 keV and
1 MeV. As can be seen, both the field and the energy dependence of
the FWHM is rather well reproduced by the simulations, but the
exact change in energy resolution is underestimated by about 1 keV
for the larger magnetic fields. Also the observed shift of the
full energy peak to lower energies (see Figs.~\ref{fig:field
spectra} and~\ref{fig:field geant}) is underestimated by the
simulations. It is impossible to say whether this discrepancy is
due to imperfections of GEANT4 when handling the altered
scattering conditions due to the presence of large magnetic
fields, or to an effect of the magnetic field on the signal
processing that is not included in the simulations, as e.g. the
charge collection process.
\\
More detailed discussions on how well GEANT4 handles these altered
scattering conditions can be found in references
\cite{Kraev:2008}\cite{Kadri:2007}\cite{Martin:2003}.
\\
\begin{table}
\begin{tabular}{ccccccc}
\hline
\hline
 & \multicolumn{2}{c}{Resolution increase} & \multicolumn{2}{c}{Resolution increase} &  \multicolumn{2}{c}{Simulated effective} \\
 B(T)& \multicolumn{2}{c}{at 485 keV (keV)} & \multicolumn{2}{c}{at 976 keV (keV)} &  \multicolumn{2}{c}{solid angle}\\
\hline
 & \underline{exp.} & \underline{sim.} & \underline{exp.} & \underline{sim.} & \underline{500 keV} & \underline{1 MeV}\\
% & \multicolumn{2}{c}{\hline} & \multicolumn{2}{c}{\hline}\\
0   & 0       & 0        & 0       & 0         & 0.03$\pi$ & 0.03$\pi$ \\
0.5 & -0.2(3) & 0.10(14) & -0.3(3) & 0.10(14)  & 0.13$\pi$ & 0.06$\pi$ \\
1   & 0.4(3)  & 0.90(14) & 0.8(3)  & 0.20(14)  & 1.4$\pi$  & 0.16$\pi$ \\
2   & 1.4(3)  & 1.73(14) & 1.9(3)  & 1.40(14)  & 1.9$\pi$  & 1.6$\pi$  \\
6   & 2.9(3)  & 1.91(14) & 3.5(3)  & 2.40(14)  & 2.0$\pi$  & 1.9$\pi$  \\
11  & 2.2(3)  & 1.90(14) & 3.6(3)  & 2.40(14)  & 2.0$\pi$  & 1.9$\pi$  \\
\hline
\hline
\end{tabular}
\caption{Comparison between the experimental and simulated
increase in energy resolution, relative to the value at 0~T, in
the presence of different magnetic fields. The experimental error
bars are extracted from the variation of the resolution at 0 T in
time. The error bars on the simulated data are coming from the
statistics of the simulation.} \label{tab:Simulation_Experiment}
\end{table}

Finally, a remark has to be made here about the long term use of
this Si PIN diode as a charged particle detector. It is well known
that the performance of semiconductor detectors degrades when
exposed to $\alpha$-radiation, which causes lattice damage in the
Si, and to cryogenic temperature cycles. We have performed the
same magnetic field tests at $\sim$10 K with a detector that had
earlier been used for $\alpha$ particle detection. This detector
performed similar as to energy resolution and scattered events in
high magnetic fields, but in fields higher then 1 T a series of
satellite peaks often appeared in the spectrum. The appearance of
these peaks did not seem to be correlated with any special event,
while they could be made to disappear by inducing a small
mechanical vibration to the system (e.g. by filling cryogenic
liquids). In order to avoid these unwanted peaks that are thought
to be related to lattice damage caused by the previously detected
$\alpha$ particles we always use new detectors in our experiments.

\section{Conclusions}

The Si PIN diode tested here showed good behavior while operating
at a temperature of $\sim$10 K in magnetic fields up to 11 T. The
response to electrons was found to change slightly, mainly due to
the influence of the magnetic field on the trajectories of the
$\beta$-particles. For mono-energetic electrons slightly (i.e.
$\sim$30\%) broadened full energy peaks were observed.
\\
The fact that this detector works well both at cryogenic
temperatures and in high magnetic fields makes it ideally suited
to be used as an energy sensitive detector for low energy
electrons and positrons as well as for $\alpha$ particles in
experimental setups where such conditions are encountered, e.g.
(cryogenic) penning traps or low temperature nuclear orientation
set-ups.
\\

We would like to thank Wim Verbruggen and Micha\"el
Vancayzeele for their assistance during the measurements. This
work was funded by the Fund for Scientific Research Flanders
(FWO), the European Union Sixth Framework Program through
RII3-EURONS (Contract no. 506065), the Grant Agency of the Czech
Republic and project GOA99-02 of the K.U.Leuven.

\end{document}